\documentclass[aps,prstab,amsmath,amssymb,twocolumn,letterpaper,%
         footinbib,bibnotes,showpacs,reprint,raggedbottom,%
         citeautoscript,floatfix,longbibliography,superscriptaddress,
         ]{revtex4-1}
\usepackage[usenames,dvipsnames,table]{xcolor}
\usepackage[%
  breaklinks,             %
  pdfpagemode   = UseNone,%
  pdfstartview  = FitH,     %
  pdfstartpage  = 1,      %
  colorlinks    = true,   %
]{hyperref}
\hypersetup{
    linkcolor=RubineRed,          
    citecolor=NavyBlue,        
    filecolor=Mulberry,      
    urlcolor=RoyalBlue           
}
\usepackage{graphicx}
\usepackage{enumerate}
\usepackage{microtype}
\usepackage{xspace}



\usepackage{xspace}
\newcommand{\supf}{\textcolor{black}{Supplementary Fig.}\xspace}

\def\srobto{SrRuO$_3$/BaTiO$_3$/SrRuO$_3$}
\def\loonno{LiOsO$_3$/NaNbO$_3$/LiOsO$_3$}

\begin{document}


\title{Polar metals as electrodes to suppress the critical-thickness limit in ferroelectric nanocapacitors}

\author{Danilo Puggioni}
\affiliation{Department of Materials Science and Engineering, 
Northwestern University, 
IL 60208-3108, USA}
\author{Gianluca Giovannetti}
\affiliation{CNR-IOM-Democritos National Simulation Centre and International School
for Advanced Studies (SISSA),
Trieste, Italy}
\author{James M.\ Rondinelli}\email{jrondinelli@northwestern.edu}
\affiliation{Department of Materials Science and Engineering, 
Northwestern University, 
IL 60208-3108, USA}

\date{\today}

\begin{abstract}
Enhancing the performance of nanoscale ferroelectric (FE) field-effect transistors and FE capacitors for memory devices and logic relies on miniaturizing the metal electrode/ferroelectric area and reducing the thickness of the insulator.
Although size reductions improve data retention, deliver lower voltage threshold switching, and increase areal density, they also degrade the functional electric polarization. There is a critical,  nanometer length $t_\mathrm{FE}^*$ below which the polarization disappears owing to depolarizing field effects. 
Here we show how to overcome the critical thickness limit imposed on ferroelectricity by utilizing electrodes formed from a novel class of materials known as polar metals. 
Electronic structure calculations on symmetric polar-metal electrode/FE capacitor structures demonstrate that electric polarizations can persist to the sub-nanometer scale with $t_\mathrm{FE}^*\rightarrow0$  
when a component of the polar axis in the electrode is perpendicular to the electrode/insulator interface.
%
Our results reveal the importance of interfacial dipolar coherency  in sustaining the polarization, which provides a platform for atomic scale structure-based design of  functions that deteriorate in reduced dimensions.
\end{abstract}

\maketitle

\section{Introduction}

%
When ferroelectric oxides are utilized in metal/oxide heterostructures and nanocapacitors, scaling of the active FE is required to improve performance \cite{Ishiwara:2004,1257065,doi:10.1080/00150199808009170}. 
Nonetheless, deleterious nanoscale effects are amplified in these geometries \cite{Junquera/Ghosez:2003,Gerra:2006, Gerra:2006v2} and act to eliminate the functional electric polarization \cite{Ishikawa:1996,Tybell:1999,Jiang:2000}.
The loss of ferroelectricity in nanoscale capacitors frequently occurs when the polarization is perpendicular to the film surface \cite{Ghosezrabe:2000, Meyer:2001, Junquera/Ghosez:2003, Kornev:2004}, \emph{i.e.}, the desired polarization direction for field-tunable devices \cite{RameshSchlom:2002}.
Bound charges 
are only partially screened at the interface, resulting in a strong depolarizing field that  
suppresses the polarization.
Indeed,  numerous reports suggest a critical thickness, $t_\mathrm{FE}^*$ below which the electric polarization disappears. 
Experimental studies on 
PbTiO$_3$ find $t_\mathrm{FE}^*\sim20$\,\AA\,  
at 300\,K \cite{Despont:2006}
whereas $t_\mathrm{FE}^*\sim4.0$\,nm,   
in Pb(Zr$_{0.2}$Ti$_{0.8}$)O$_3$ films \cite{Tybell:1999}. 
Furthermore, first-principles density functional theory (DFT) calculations predict $t_\mathrm{FE}^*\sim2.4$\,nm in single-domain BaTiO$_3$ films between SrRuO$_3$ electrodes \cite{Junquera/Ghosez:2003}, which reduces to $t_\mathrm{FE}^*\sim1.0$\,nm  after accounting for ionic relaxation in the electrodes \cite{Gerra:2006, Gerra:2006v2}. 
With this limitation, ferroelectric based devices are unable to meet the continuous scaling changes demanded by higher density data storage technologies.
%

Although proposals to overcome the problem exist, a general solution remains elusive. 
Most approaches focus on tuning the stability of the FE state. 
Epitaxial strain engineering has been proposed; nonetheless, this strategy extends to a limited number of oxides, requires complex processing steps, and is limited by available commercial substrates. For example, although it is predicted that $t_\mathrm{FE}^*\rightarrow0$ in 
thin films of the incipient ferroelectric BaZrO$_3$, a large epitaxial compressive strain of 4.25\% is required \cite{Zhang:2014}, 
which would produce deleterious misfit dislocations. 
Integration with silicon is likely to also lead to uncontrolled interface states \cite{Hirth:2008}.
%
%
Furthermore, at this level of strain $t_\mathrm{FE}^*$ is still finite for BaTiO$_3$ (Ref.\ \onlinecite{Zhang:2014}). 

Alternative solutions change the type of ferroelectric and the active inversion symmetry lifting mechanism. Ab-initio calculations find that $t_\mathrm{FE}^*\!\rightarrow\!0$ using so-called hyperferroelectrics \cite{Garrity:2014}, which have a persistent polarization different from proper FE oxides,  or improper ferroelectrics \cite{Stengel:2012,PhysRevLett.102.107601}. However, hyperferroelectric bulk  materials (and thin films) remain to be synthesized  \cite{Bennett:2012}.
%
Another route relies on creating an enhanced interfacial FE state by controlling the covalency of the metal-oxygen bond at the heterointerface \cite{Stengel/Spaldin:2006,Stengel/Vanderbilt/Spaldin:2009a,Caimeng:2011,ADFM:ADFM201500371}.
%

\begin{figure}[t]
\includegraphics[width=0.95\columnwidth]{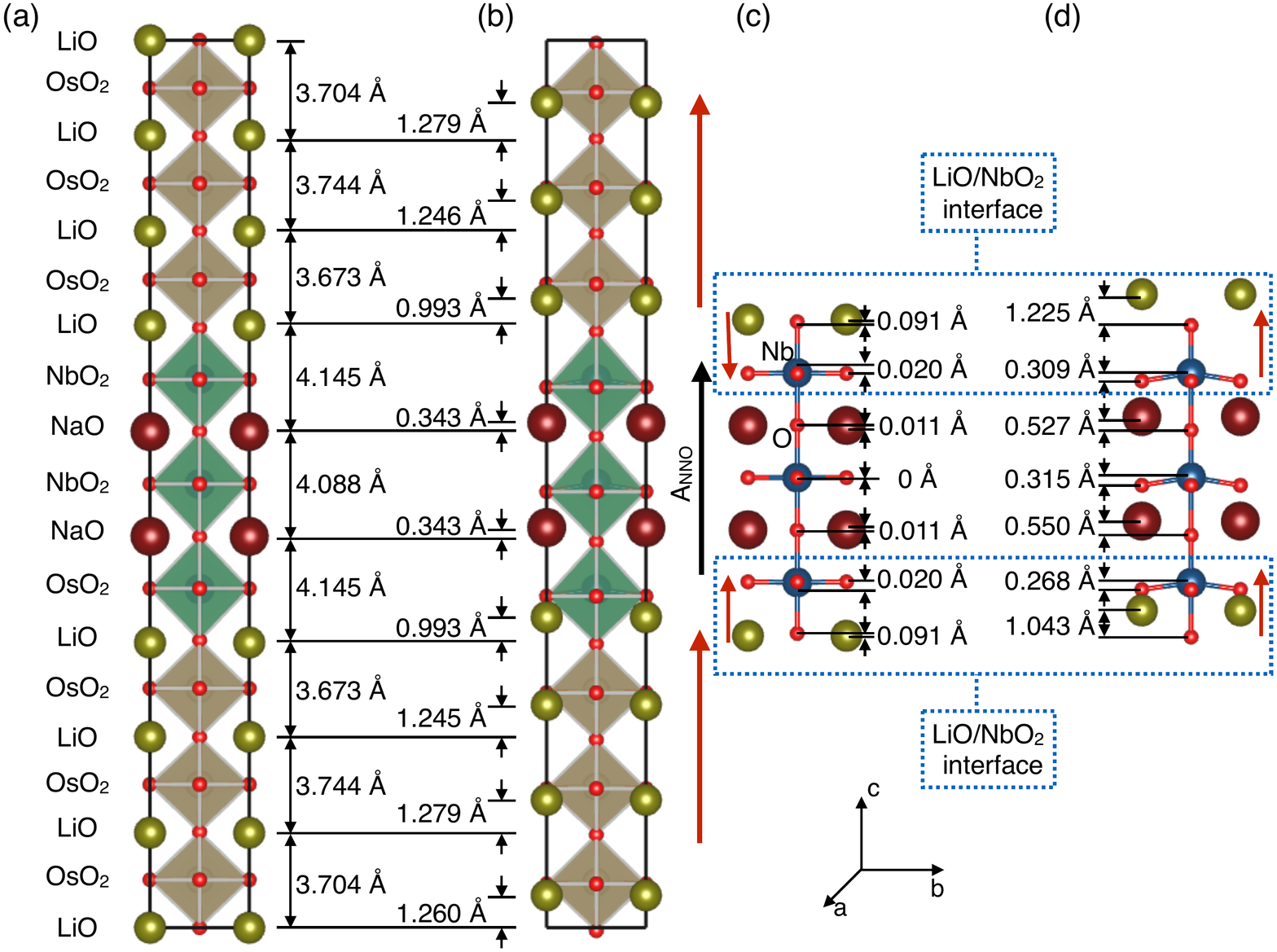}
\caption{{
Symmetric nanocapacitor consisting of polar-metal electrodes and a  ferroelectric oxide.} \textbf{a},  The  centrosymmetric nanocapacitor with insulating NaNbO$_3$ ($m$=2) between LiOsO$_3$ electrodes ($n$=6). \textbf{b}, The equilibrium structure of the ferroelectric capacitor  [LiO-(OsO$_2$-LiO)$_6$/NbO$_2$-(NaO-NbO$_2$)$_2$]. The direction of the polar displacements in the electrodes and the ferroelectric film ($\mathcal{A}_\mathrm{NNO}$) are indicated with arrows. Magnification of the LiO/NbO$_2$ interface of the  \textbf{c}, paraelectric aristotype and  \textbf{d}, the ground state structure.  
The FE behavior at the interface is due to the Li, Nb, and O displacements. The direction of the polar displacements at the interfaces for the paraelectric and the ground state structure is indicated with red arrows.}
\label{fig1}
\end{figure}

Here we examine the critical thickness for ferroelectricity in nanocapacitors consisting of \emph{polar-metal} electrodes and conventional ferroelectric oxides under short-circuit boundary conditions using 
 DFT calculations.  
Recently, polar metals have garnered  considerable interest \cite{Shi:2013,Kim:2016} because they 
exhibit simultaneously inversion-lifting displacements and metallicity.
In these compounds, the polar displacements are weakly coupled to the states at the Fermi level, which makes possible the coexistence of a polar structure and metallicity \cite{Puggioni:2013}. 
%
Our main finding is that polar-metal  electrodes suppress the critical thickness limit through interfacial polar displacements, which stabilize the ferroelectric (polarized) state; this geometric effect does not rely on interfacial bond chemistry or `perfect' screening of the depolarizating field,  but rather results from the intrinsic broken parity present in the electrode.



\section{Methods}
%
We perform first-principles DFT calculations
within the local-density approximation (LDA) and hybrid functional (HSE06, Ref.~\onlinecite{Heyd2003,Heyd2006})
as implemented in the Vienna
{\it Ab initio} Simulation Package (VASP) 
\cite{Kresse/Furthmuller:1996b} 
with the projector augmented wave (PAW) approach \cite{Blochl/Jepsen/Andersen:1994} 
to treat the core and valence electrons using 
the following  electronic configurations:  
$1s^2 2s^2$ (Li),
$5p^{6} 6s^2 5d^6$ (Os), 
$2s^2 2p^4$ (O),
$5s^2 5p ^6 6s^2$ (Ba), 
$3d^2 4s ^2$ (Ti),
$4s^2 4p ^6 5s^2$ (Sr), 
$4d^7 5s ^1$ (Ru), 
$2p^6 3s ^1$ (Na), 
$4p^6 4d ^4 5s^1$ (Nb).
The Brillouin zone integrations are performed 
with a  $13\times13\times1$
Monkhorst-Pack $k$-point mesh \cite{Monkhorst/Pack:1976} 
and a 600~eV plane wave cutoff for the 
\loonno\ 
and 
\srobto\ 
capacitor structures.
We relax the atomic positions (force tolerance less than 0.1\,meV~\AA$^{-1}$) using Gaussian smearing (20\,meV width).
Below 150\,K NaNbO$_3$ and LiOsO$_3$ are isostructural with rhombohedral space group $R3c$ and pseudocubic lattice parameters of 3.907~{\AA} (Ref.~\onlinecite{Zeitschrift}) and 3.650~{\AA} (Ref.~\onlinecite{Shi:2013}), respectively.  
%
%
Owing to the large lattice mismatch between the two compounds, we simulate a symmetric ferroelectric capacitor structure with an LiO/NbO$_2$ interfacial termination, 
shown in \autoref{fig1}a, under an epitaxial  constraint that would be imposed by a (La$_{0.3}$Sr$_{0.7}$)(Al$_{0.65}$Ta$_{0.35}$)O$_3$ substrate\cite{Liu:2013} and we relax the out-of-plane lattice parameter. 
This results in a compressive strain of $\sim$1\% for NaNbO$_3$ and a tensile strain of $\sim$6\% for LiOsO$_3$. Note that at the bulk level, we find that a tensile strain $\gtrsim$6\% suppresses the polar instability along the [001]-pseudocubic direction of LiOsO$_3$. Moreover, we  selected NaNbO$_3$ for the capacitor structures because to eliminate any charge transfer due to `polar catastrophe/charge mismatch' physics as the interface: [LiO]$^{1-}$, [NaO]$^{1-}$, [NbO2]$^{1+}$,  and [OsO2]$^{1+}$.

For the two ferroelectric capacitors, we adopt the layered-oxide notation used in Ref.~\onlinecite{Junquera/Ghosez:2003}, that is
\begin{itemize}\itemsep=0em
\item {[LiO-(OsO$_2$-LiO)$_n$/NbO$_2$-(NaO-NbO$_2$)$_m$]}  and  \item {[SrO-(RuO$_2$-SrO)$_n$/TiO$_2$-(BaO-TiO$_2$)$_m$]}
\end{itemize}
to clearly demarcate the interface composition in the \loonno\ and \srobto\ capacitors, respectively.
We use a LiO/NbO$_2$ electrode/ferroelectric interface for the \loonno\  capacitor and a SrO/TiO$_2$  interface termination for the \srobto\  capacitor. 
For both 
ferroelectric capacitors, we constrained the number of 5-atom perovskite units of the electrode at $n=6$ 
to ensure a thickness large enough to avoid interaction between the two interfaces, and $m$ ranged from 1 to 3.  
The periodic boundary conditions naturally impose the required short-circuit condition between the electrodes.
Note that for \srobto\ capacitors, our geometry differs only slightly from that used by Junquera and Ghosez  \cite{Junquera/Ghosez:2003}, whereby the thickness of our electrode is greater.

The group theoretical analysis was aided by the \textsc{isodistort}  software \cite{Isodisplace:2006}. 
It is used to evaluate the geometric-induced inversion symmetry-breaking displacements of the $P4mm$ structure with respect the $P4/mmm$ phase, reducing the polar structure into a set of symmetry-adapted modes associated with different irreducible representations of the $P4/mmm$ phase. The ``robust'' algorithm was used to match an atom in the undistorted structure to every atom in the distorted structure separated by a threshold distance less than 3\,\AA.

\section{Results and Discussion}
%
%
The first ferroelectric nanocapacitor we focus on consists of ferroelectric NaNbO$_3$ of varying thickness $m$ confined between electrodes of the experimentally known polar metal LiOsO$_3$ (see  \autoref{fig1}, $m=2$) \cite{Shi:2013,PhysRevB.90.195113}. 
We adopt the layered-oxide notation used in Ref.~\onlinecite{Junquera/Ghosez:2003}, that is
[LiO-(OsO$_2$-LiO)$_n$/NbO$_2$-(NaO-NbO$_2$)$_m$]
to clearly demarcate the interface composition (see Methods).
We create two symmetric nanocapacitors with a polar and paraelectric configuration for both LiOsO$_3$ and NaNbO$_3$, respectively. 
We then relax the out-of-plane lattice parameter and the atomics positions of the nanocapacitors for  
$m$=1, 2, and 3. 
The lowest energy heterostructures are polar with space group $P4mm$ and exhibit large Li ions displacements along the [001]-pseudocubic direction (\autoref{fig1}b). No zone-center dynamical instabilities are found in these heterostructures. 
Note that structures with an initial paraelectric configuration relax into a centrosymmetric structure (space group $P4/mmm$) with Li atoms displaying large antipolar displacements in LiOsO$_3$ (\supf1) that decrease towards the interface (\autoref{fig1}c).

We use representation theory analysis to examine the inversion lifting distortions (see Methods), and find that the distortion vector corresponds to the irreps $\Gamma_1^+$ and $\Gamma_3^-$. The irrep $\Gamma_1^+$ reduces the antipolar displacements in 
LiOsO$_3$ resulting in the centrosymmetric $P4/mmm$ structure depicted in \autoref{fig1}a. Differently, the irrep $\Gamma_3^-$ is a polar mode which involves mainly Li ion displacements---the maximum amplitudes being $\sim$1.3~\AA\ in LiOsO$_3$ with decreasing amplitude towards the LiO/NbO$_2$ interface (\autoref{fig1}b). It also consists of polar displacements of all ions in the dielectric NaNbO$_3$ layers with the Nb ions off-centering the most (\autoref{fig1}d) 

%
For all thicknesses, the NaNbO$_3$ thin film maintains a ferroelectric ground state characterized by both Nb and Na displacements (\autoref{fig1}b).
%
A linear polarization-displacement model using  the Born effective charge from Ref.~\onlinecite{PhysRevLett.72.3618} 
results in a 0.86\,C\,m$^{-2}$ polarization for NaNbO$_3$ in $m=2$.
%
An analysis of the differential ionic relaxations in the heterostructure reveals polar displacements at the LiO/NbO$_2$ interfaces---an interfacial ferroelectricity---which are a consequence of the polar metal used as an electrode. 
%
This produces an enhanced polarization in the ferroelectric compared to that calculated using the aforementioned procedure in the experimental $R3c$ structure (0.59\,C\,m$^{-2}$) \cite{PhysRevB.76.024110}.
%
In particular, although the two interfaces of the paraelectric structures exhibit \emph{antiparallel} polar displacements (\autoref{fig1}c and  \supf1), the interfaces of the polar ground state structures have \emph{parallel} polar displacements as show in \autoref{fig1}d.

\begin{figure}
\includegraphics[width=0.9\columnwidth]{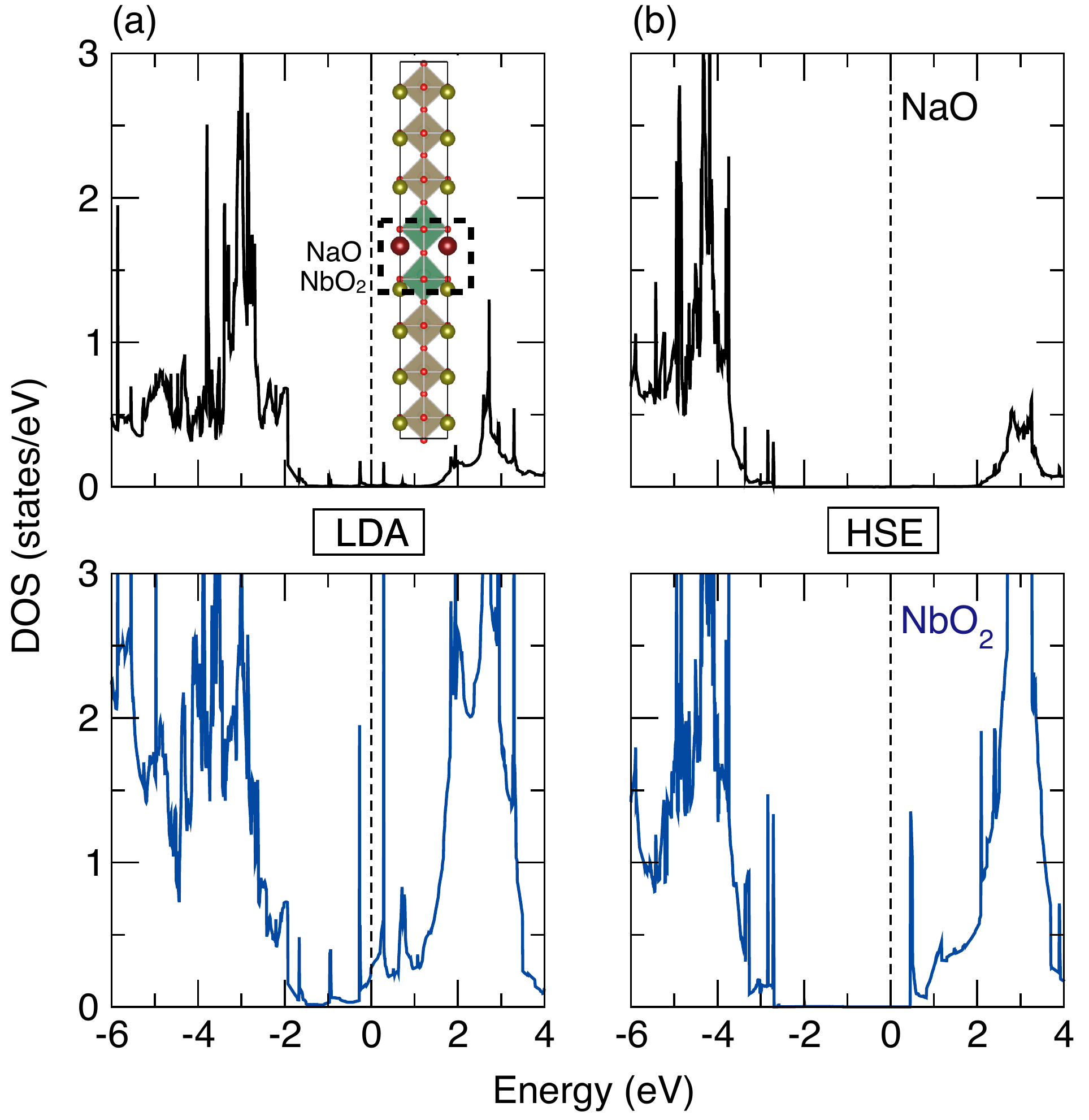}
\caption{Projected densities-of-states (DOS) for the NaNbO$_3$ layer in the \loonno\ ($m$=1) nanocapacitor within  \textbf{a}, LDA and \textbf{b}, HSE06. The inset shows the  \loonno\ ($m$=1) nanocapacitor.}
\label{fig_dos}
\end{figure}

\autoref{fig_dos} shows the electronic properties for the NaNbO$_3$ layer in the \loonno\ ($m$=1) nanocapacitor.  The LDA functional predicts that the NaNbO$_3$ film is metallic, rendering the ferroelectric capacitor unusable (\autoref{fig_dos}a). This behavior is artificial and due to the tendency of the LDA functional to underestimate the band gap in insulating compound \cite{Perdewgap:1985,stengelpuente:2011}. We solve this pathological problem for DFT by using a more sophisticated functional  which includes a fraction of exact exchange (HSE06). \autoref{fig_dos}b shows that the hybrid functional fully opens the gap between the O $2p$ and Nb $4d$ states. Moreover, we find that the HSE06-relaxed structures exhibit displacements similar to those obtained from the LDA functional; importantly, polar displacements at the LiO/NbO$_2$ interfaces. This result supports the conclusion that the interfacial ferroelectricity is induced by the polar crystal structure of the metallic electrode and not due to spurious shorting of the capacitor. In the remainder of this paper, we report results obtained using LDA owing to the similar crystal structure obtained with HSE06 functional.

\begin{figure}[t]
\centering
\includegraphics[width=0.9\columnwidth]{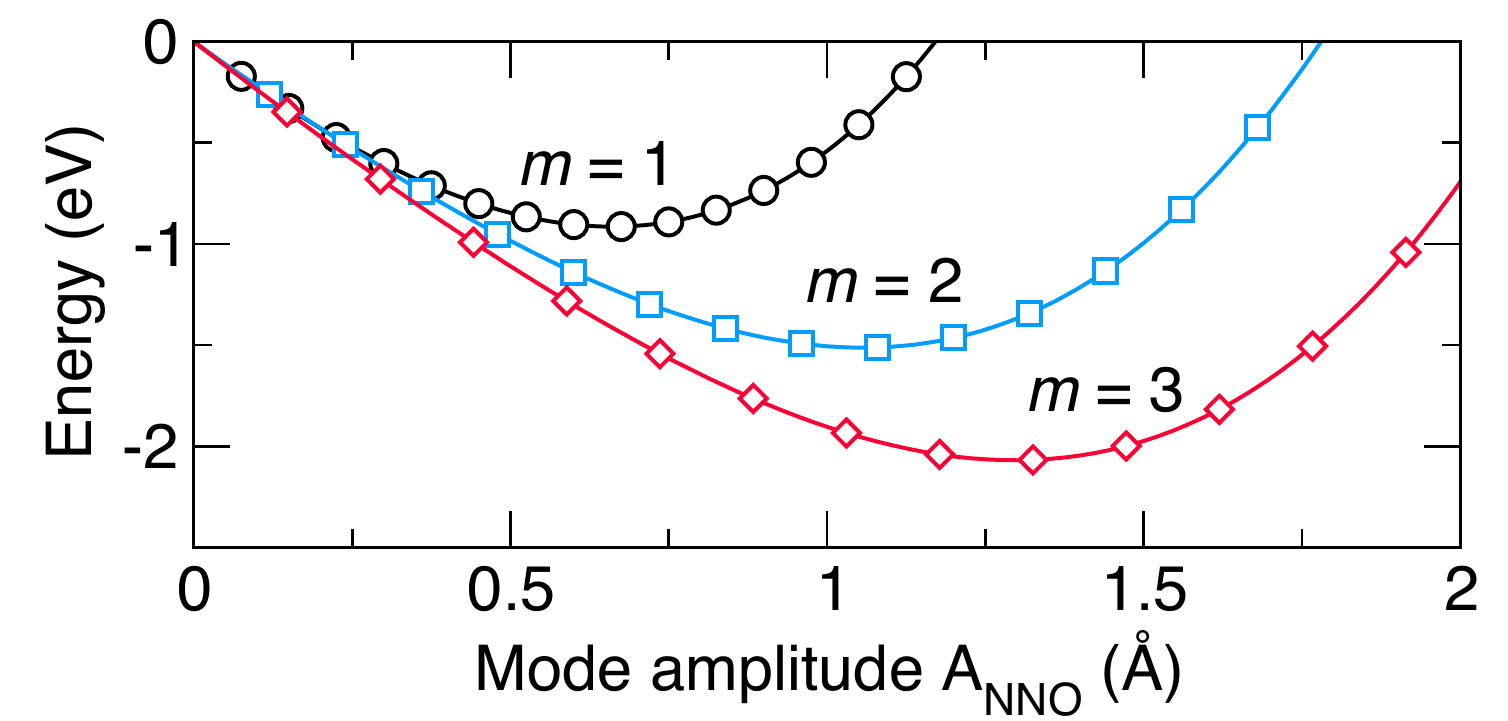}
\caption{
Energetic gain with increasing polar mode amplitude $\mathcal{A}_\mathrm{NNO}$ in the NaNbO$_3$ film for different thicknesses: $m$=1  (circles), $m$=2 (squares) and  $m$=3 (diamonds). A finite value of  $\mathcal{A}_\mathrm{NNO}$ leads to an energetic gain for all $m$, indicating the disappearance of the critical thickness. Note that the energy of the polar ground state structure with $\mathcal{A}_\mathrm{NNO}$=0~{\AA} is taken as reference for each value of $m$. As expected the energy gain increases as the thickness of the ferroelectric film increases owing to the reduced effects from the depolarizing field, which dominates thinner films.}
%
\label{fig2}
\end{figure}

\autoref{fig2} shows the evolution of the total energy of each capacitor 
with mode amplitude $\mathcal{A}_\mathrm{NNO}$, which describes the atomic displacements involved in the soft mode of the NaNbO$_3$ film. 
As expected the largest energy gain occurs when the thickness of the ferroelectric film increases. Note that the shape of the energy surface does not exhibit the characteristic double well behavior, because in our calculations we 
fix the polar displacements in the metallic electrodes 
and only change the amplitude of the polar displacements in NaNbO$_3$.
%
%
Independent of the NaNbO$_3$ film thickness, the energy is minimized for the ferroelectric ground state ($\mathcal{A}_\mathrm{NNO}\neq0$), indicating that an ideal ferroelectric capacitor can be reduced to an ultrathin (single unit cell) size, i.e., $t_\mathrm{FE}^*\rightarrow0$. 
%

The disappearance of the critical-thickness limit to ferroelectricity is the result of the parallel polar displacements present at the electrode/dielectric interfaces (\autoref{fig1}b).
As proposed in Ref.\ \onlinecite{Stengel/Vanderbilt/Spaldin:2009a}, the ferroelectric state in ultrathin-film devices depends crucially on the nature of the chemical bonds at the metal/oxide interface. 
Here, this interfacial bonding occurs and is an immediate consequence of the structure of the polar-metal electrodes. The enhanced and parallel interfacial polar displacements ``imprint'' and lead to an overall enhancement of the ferroelectric instability of the film, which we assess further below. We stress that differently from Ref.\ \onlinecite{Stengel/Vanderbilt/Spaldin:2009a}, the interfacial dipole distortions are due to a geometrical mechanism driven by the polar structure of the metallic electrode and not due to the stiffness of the electrode--oxide bonds.
\begin{figure}
\includegraphics[width=\columnwidth]{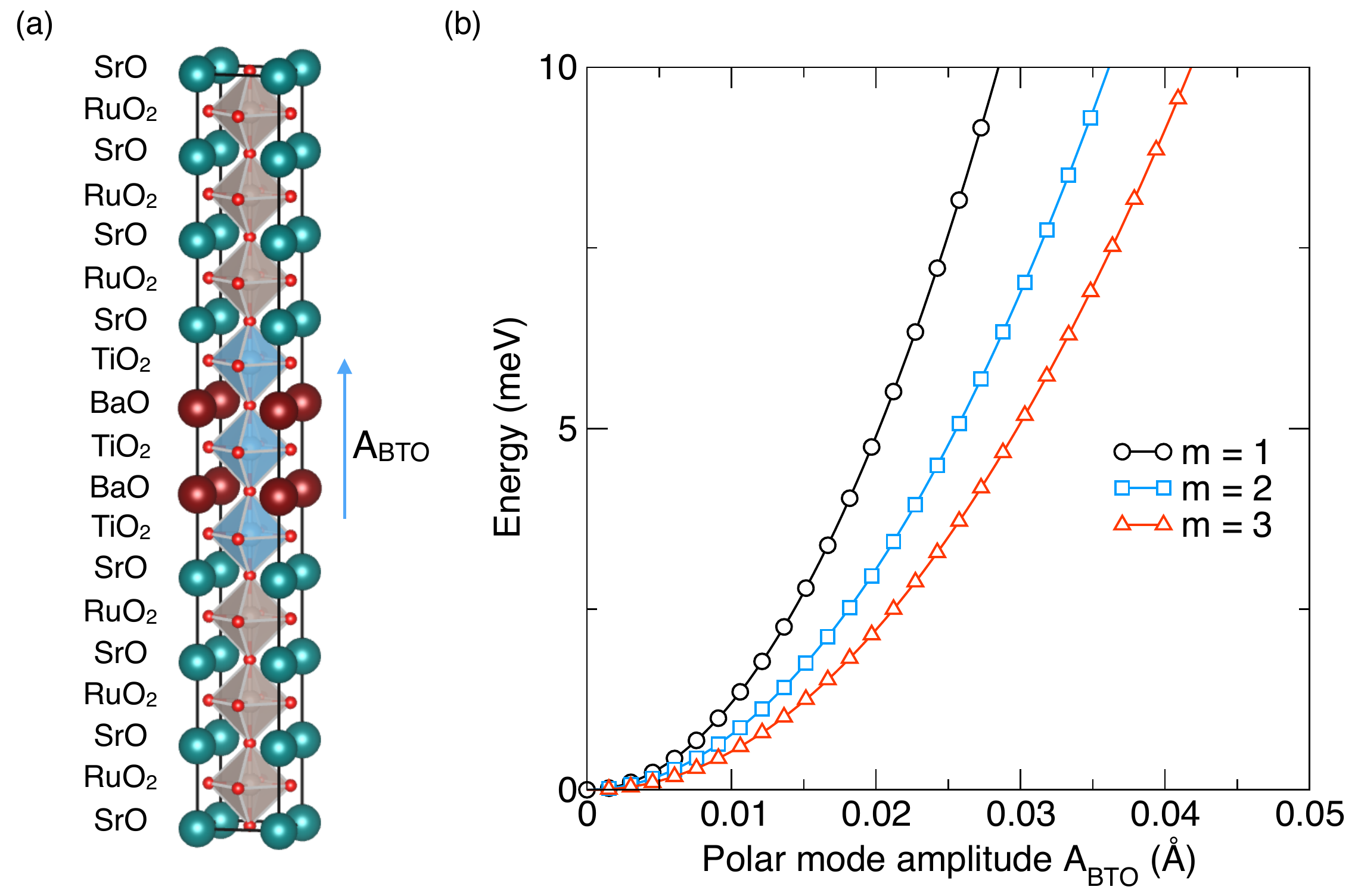}
\caption{{Typical behavior of a nanoscale ferroelectric capacitor below the critical thickness.} \textbf{a}, Nanocapacitor consisting of a BaTiO$_3$ film ($m$=2)  between  centrosymmetric SrRuO$_3$ ($n$=6) electrodes. The arrow indicates the direction of the polar displacements in BaTiO$_3$. 
%
%
\textbf{b}, Energy as a function of the polar mode amplitude $\mathcal{A}_\mathrm{BTO}$ in BaTiO$_3$. The energy increases for all  thicknesses $m$ analyzed. 
}
\label{fig3}
\end{figure}
%

%

We next examine a \srobto\ capacitor (\autoref{fig3}a), which as before may be written as 
{[SrO-(RuO$_2$-SrO)$_n$/TiO$_2$-(BaO-TiO$_2$)$_m$]}
to reveal the layered monoxide planes in the structures.
We focus on the SrO/TiO$_2$ interface geometry to demonstrate the generality of this solution to the critical thickness problem.
We constrain the in-plane lattice parameters to that of SrTiO$_3$ (3.905\, \AA) and 
we relax the out-of-plane lattice parameter and the atomic positions (see Methods), examining 
capacitors with $m$=1, 2, and 3 that are well below the reported $m$=7 critical thickness \cite{Junquera/Ghosez:2003}. 
With nonpolar SrRuO$_3$ electrodes the paraelectric configuration is  energetically more stable than the ferroelectric configuration for all BaTiO$_3$ film thicknesses  studied. 
This is confirmed by the increase in total energy as a function of the polar mode amplitude $\mathcal{A}_\mathrm{BTO}$ (\autoref{fig3}b). 
Indeed, the use of a centrosymmetric metal for the thinnest ferroelectric film results in an antisymmetric poling effect of the two interfaces, which forbids the possibility of a ferroelectric displacement \cite{Gerra:2006}. 

\begin{figure}
\centering
\includegraphics[width=0.9\columnwidth]{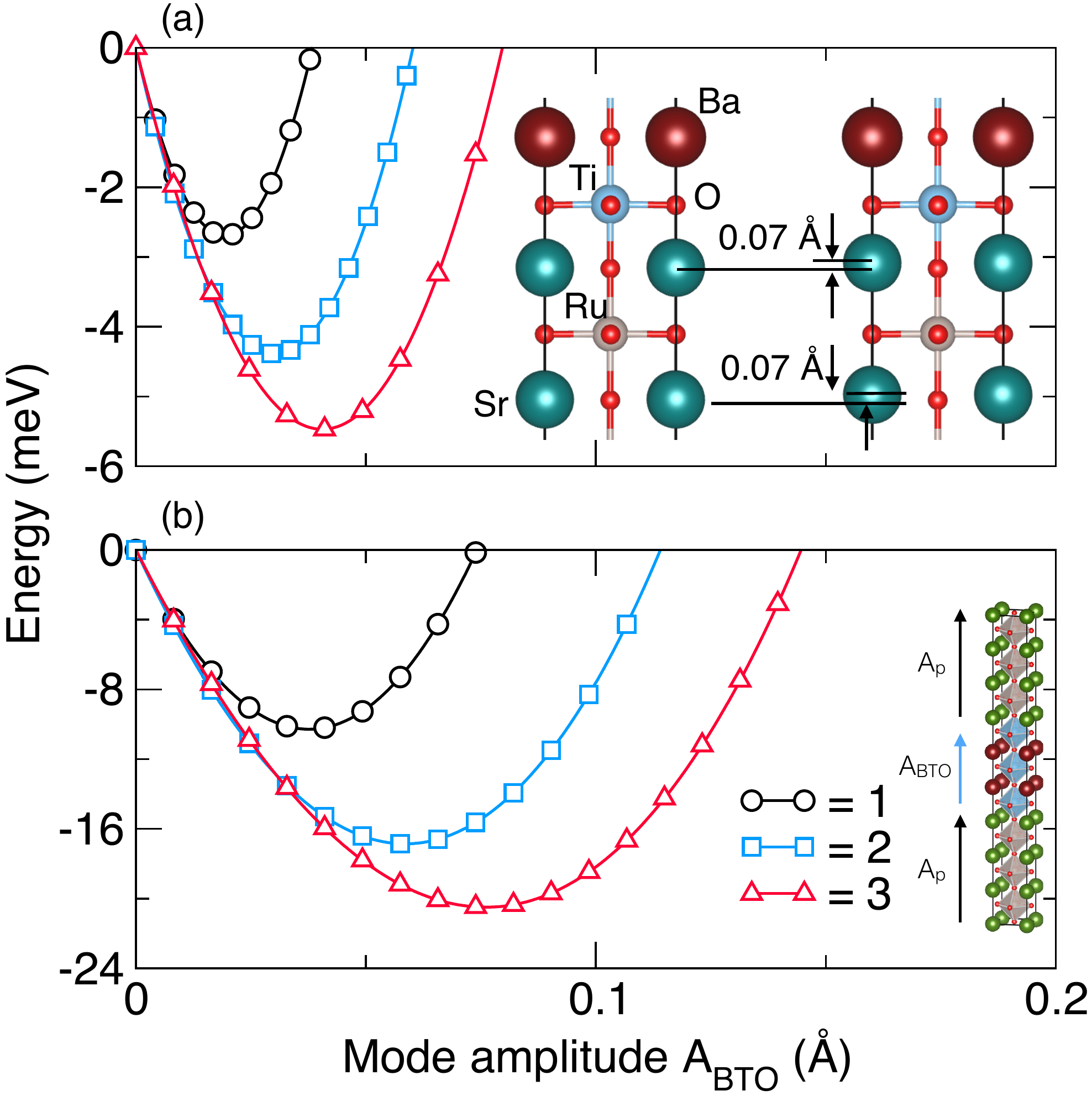}
\caption{{Energetic landscape of a BaTiO$_3$ nanocapacitor with ``polar'' SrRuO$_3$ electrodes.} Energetic gain for different dielectric thicknesses: $m$=1  (circles), $m$=2 (squares) and  $m$=3 (diamonds). The Sr displacements given with respect to the centrosymmetric ground state are  fixed to \textbf{a}, 0.07~{\AA} (see inset) and \textbf{b},  0.14~{\AA}.
%
%
The energy of the polar  structure with $\mathcal{A}_\mathrm{BTO}$=0~{\AA} is taken as reference for each $m$.}
\label{fig4}
\end{figure}

Now we perform a computational experiment whereby we transmute centric SrRuO$_3$ into a hypothetical polar metal by following the design rules for noncentrosymmetric metals introduced in Ref.~\onlinecite{Puggioni:2013}. We do this by imposing a polar distortion in SrRuO$_3$,  which involves only the Sr atoms, as the orbital character at the Fermi level has a negligible  contribution from these atoms, with 
parallel polar displacements 
at the SrO/TiO$_2$ interfaces as suggested by our \loonno\ capacitor results. We point out that bulk SrRuO$_3$ does not exhibit polar distortions and here we make it artificially polar to isolate the interfacial geometric effect independent of chemistry with respect to the model with centrosymmetric electrodes.


%

In \autoref{fig4}a we show the energy evolution of this hypothetical capacitor 
as a function of the mode amplitude $\mathcal{A}_\mathrm{BTO}$, with  parallel polar Sr displacements imposed uniformly at 0.07~{\AA} with respect to the centrosymmetric structure (see inset). 
%
%
In contrast to \autoref{fig3}a, the ferroelectric state of BaTiO$_3$ is more stable than the paraelectric geometry for all thicknesses $m$=1, 2, and 3 as indicated by the energy gain for $\mathcal{A}_\mathrm{BTO}\ne0$. These results indicate that $t_\mathrm{FE}^*\rightarrow0$
 in the BaTiO$_3$ film between the polar-metal SrRuO$_3$.
%

The energy gain is strongly influenced by the polar displacements of the Sr atoms. 
In particular, by doubling the amplitude of the polar displacements of the Sr atoms, from 0.07~{\AA} to 0.14~{\AA}  for the case $m$=1, we find that the energy gain increases from $\sim$3~meV to $\sim$10~meV 
(\autoref{fig4}b). 
%
Comparing \autoref{fig4}a and \autoref{fig4}b, we find a shift in the critical mode amplitude to larger values, 
which suggests that the device containing a polar-metal electrode with larger Sr displacements displays a larger ferroelectric polarization. 
Indeed, when we consider the other limit by fully removing the  polar displacements at the SrO/TiO$_2$ interface (setting them to 0\,\AA), the energy landscape presented in \autoref{fig3}b is restored. 
Note that for the disappearance of the critical thickness, it is necessary that the polar direction (or a component of it) in the electrodes, and therefore that of the interfacial dipole, coincides with the direction of polarization of the ferroelectric film (\autoref{fig4}b, inset).

\begin{figure}
\includegraphics[width=\columnwidth]{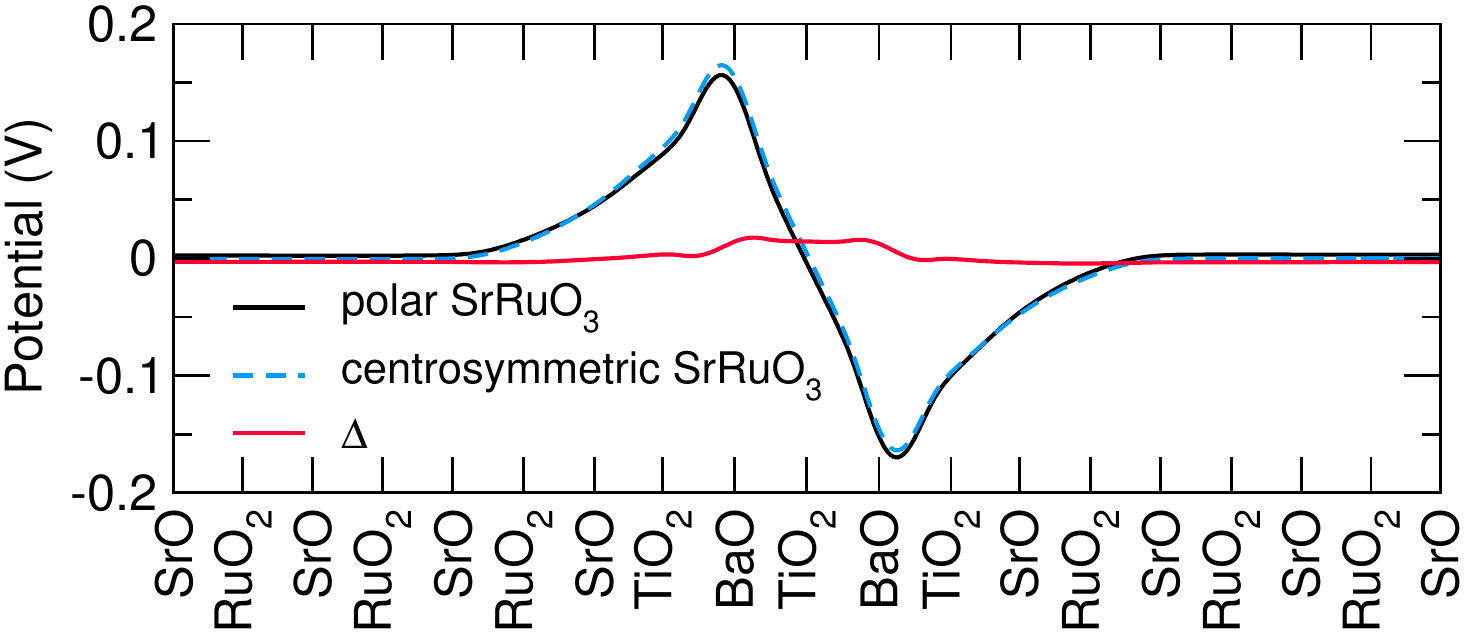}
\caption{
The planar and macroscopically averaged electrostatic potential of the BaTiO$_3$  ferrocapacitor ($m$=2) along [001] between 
polar (solid line) and centrosymmetric  (broken line) SrRuO$_3$ electrodes. The difference, $\Delta$, is also shown. 
The amplitude of the polar distortion in the BaTiO$_3$ film is fixed to $\sim0.6$~\AA, which corresponds to the minimum in \autoref{fig4}b. The configuration with paraelectric BaTiO$_3$ is used as reference.}
\label{fig6}
\end{figure}

One could argue that the findings here are a result of the polar metals better screening the ferroelectrics polarization; however, this is not the case. Indeed the electrostatic for the BaTiO$_3$ nanocapacitor with ``polar''  and nonpolar SrRuO$_3$ electrodes are almost the same (\autoref{fig6}). 
Moreover, polar metals typically have longer screening lengths than conventional metals \cite{Giovannetti_etal:2016}.
This result further confirms the role of interfacial geometric effects induced by the polar structure of the metallic electrode in controlling 
the critical thickness.

Lastly, we discuss how the proposed device can be switched. 
Polar metals are not ferroelectrics. Indeed, the application of an electric field cannot switch the polar distortion in the metal because the free electrons will screen the electric field.
However, it has been show that the polar distortion in the metal can be switched 
by applying an electric field to a superlattice composed of  an insulating ferroelectric material and a polar metal 
by coupling to the ferroelectric polarization \cite{Xiang:2014}.  
Similarly, when an electric field is applied to the aforementioned nanocapacitors, the polar distortion in the ferroelectric thin film should align along the direction of the electric field; then because of the interfacial coupling between the polar metallic electrode and the ferroelectric film \cite{Xiang:2014}, the polar displacements in the polar metal and the interfacial polar displacements should follow. Note that the geometric configuration required to sustain ferroelectricity, \emph{i.e.}, $t_\mathrm{FE}^*\rightarrow0$, 
is preserved in the switching mechanism. Alternative approaches have also been applied to degenerately doped ferroelectrics.\cite{Nukala:2017}

\section{Conclusion}

In summary, we proposed a ferroelectric capacitor where the conventional metallic electrodes are replaced by \emph{noncentrosymmetric} metallic electrodes. 
We showed that the polar displacements in the noncentrosymmetric metallic electrodes induce interfacial ferroelectricity, which supports a polar instability in the ferroelectric film  regardless of the dielectric thickness.
%
%
Although we utilized LiOsO$_3$ herein for simplicity, we point out that our result is completely general and the same conclusions may be achieved using other noncentrosymmetric metals as electrodes with our described geometric constrations  (see Ref.~\onlinecite{Benedek:2016} for a list of materials). 
These polar-metal based nanoscale capacitors maintain the functionality of the ferroelectric film independent of the degree of miniaturization and could lead to device architectures with improved scalability.
%

\begin{acknowledgements}
D.P.\ and J.M.R.\ acknowledge the Army Research Office under Grant No.\ W911NF-15-1-0017 for financial support and the DOD-HPCMP for computational resources. 
\end{acknowledgements}

\bibliography{NCSM_Electrode_NNano_GG}

\begin{thebibliography}{44}%
\makeatletter
\providecommand \@ifxundefined [1]{%
 \@ifx{#1\undefined}
}%
\providecommand \@ifnum [1]{%
 \ifnum #1\expandafter \@firstoftwo
 \else \expandafter \@secondoftwo
 \fi
}%
\providecommand \@ifx [1]{%
 \ifx #1\expandafter \@firstoftwo
 \else \expandafter \@secondoftwo
 \fi
}%
\providecommand \natexlab [1]{#1}%
\providecommand \enquote  [1]{``#1''}%
\providecommand \bibnamefont  [1]{#1}%
\providecommand \bibfnamefont [1]{#1}%
\providecommand \citenamefont [1]{#1}%
\providecommand \href@noop [0]{\@secondoftwo}%
\providecommand \href [0]{\begingroup \@sanitize@url \@href}%
\providecommand \@href[1]{\@@startlink{#1}\@@href}%
\providecommand \@@href[1]{\endgroup#1\@@endlink}%
\providecommand \@sanitize@url [0]{\catcode `\\12\catcode `\$12\catcode
  `\&12\catcode `\#12\catcode `\^12\catcode `\_12\catcode `\%12\relax}%
\providecommand \@@startlink[1]{}%
\providecommand \@@endlink[0]{}%
\providecommand \url  [0]{\begingroup\@sanitize@url \@url }%
\providecommand \@url [1]{\endgroup\@href {#1}{\urlprefix }}%
\providecommand \urlprefix  [0]{URL }%
\providecommand \Eprint [0]{\href }%
\providecommand \doibase [0]{http://dx.doi.org/}%
\providecommand \selectlanguage [0]{\@gobble}%
\providecommand \bibinfo  [0]{\@secondoftwo}%
\providecommand \bibfield  [0]{\@secondoftwo}%
\providecommand \translation [1]{[#1]}%
\providecommand \BibitemOpen [0]{}%
\providecommand \bibitemStop [0]{}%
\providecommand \bibitemNoStop [0]{.\EOS\space}%
\providecommand \EOS [0]{\spacefactor3000\relax}%
\providecommand \BibitemShut  [1]{\csname bibitem#1\endcsname}%
\let\auto@bib@innerbib\@empty
\bibitem [{\citenamefont {Ishiwara}\ \emph {et~al.}(2004)\citenamefont
  {Ishiwara}, \citenamefont {Okuyama},\ and\ \citenamefont
  {Arimoto}}]{Ishiwara:2004}%
  \BibitemOpen
  \bibfield  {author} {\bibinfo {author} {\bibfnamefont {H.}~\bibnamefont
  {Ishiwara}}, \bibinfo {author} {\bibfnamefont {M.}~\bibnamefont {Okuyama}}, \
  and\ \bibinfo {author} {\bibfnamefont {Y.}~\bibnamefont {Arimoto}},\
  }\href@noop {} {\emph {\bibinfo {title} {Ferroelectric Random Access
  Memories: Fundamentals and Applications}}}\ (\bibinfo  {publisher}
  {Springer},\ \bibinfo {address} {Berlin, Germany},\ \bibinfo {year}
  {2004})\BibitemShut {NoStop}%
\bibitem [{\citenamefont {Muller}\ \emph {et~al.}(2003)\citenamefont {Muller},
  \citenamefont {Nagel}, \citenamefont {Pinnow},\ and\ \citenamefont
  {Rohr}}]{1257065}%
  \BibitemOpen
  \bibfield  {author} {\bibinfo {author} {\bibfnamefont {G.}~\bibnamefont
  {Muller}}, \bibinfo {author} {\bibfnamefont {N.}~\bibnamefont {Nagel}},
  \bibinfo {author} {\bibfnamefont {C.~U.}\ \bibnamefont {Pinnow}}, \ and\
  \bibinfo {author} {\bibfnamefont {T.}~\bibnamefont {Rohr}},\ }\bibfield
  {title} {\enquote {\bibinfo {title} {Emerging non-volatile memory
  technologies},}\ }in\ \href {\doibase 10.1109/ESSCIRC.2003.1257065} {\emph
  {\bibinfo {booktitle} {Solid-State Circuits Conference, 2003. ESSCIRC '03.
  Proceedings of the 29th European}}}\ (\bibinfo {year} {2003})\ pp.\ \bibinfo
  {pages} {37--44}\BibitemShut {NoStop}%
\bibitem [{\citenamefont {Scott}(1998)}]{doi:10.1080/00150199808009170}%
  \BibitemOpen
  \bibfield  {author} {\bibinfo {author} {\bibfnamefont {J.~F.}\ \bibnamefont
  {Scott}},\ }\bibfield  {title} {\enquote {\bibinfo {title} {Future issues in
  ferroelectric miniaturization},}\ }\href {\doibase 10.1080/00150199808009170}
  {\bibfield  {journal} {\bibinfo  {journal} {Ferroelectrics}\ }\textbf
  {\bibinfo {volume} {206}},\ \bibinfo {pages} {365--379} (\bibinfo {year}
  {1998})},\ \Eprint
  {http://arxiv.org/abs/http://dx.doi.org/10.1080/00150199808009170}
  {http://dx.doi.org/10.1080/00150199808009170} \BibitemShut {NoStop}%
\bibitem [{\citenamefont {Junquera}\ and\ \citenamefont
  {Ghosez}(2003)}]{Junquera/Ghosez:2003}%
  \BibitemOpen
  \bibfield  {author} {\bibinfo {author} {\bibfnamefont {Javier}\ \bibnamefont
  {Junquera}}\ and\ \bibinfo {author} {\bibfnamefont {Philippe}\ \bibnamefont
  {Ghosez}},\ }\bibfield  {title} {\enquote {\bibinfo {title} {Critical
  thickness for ferroelectricity in perovskite ultrathin films},}\ }\href@noop
  {} {\bibfield  {journal} {\bibinfo  {journal} {Nature}\ }\textbf {\bibinfo
  {volume} {422}},\ \bibinfo {pages} {506--509} (\bibinfo {year}
  {2003})}\BibitemShut {NoStop}%
\bibitem [{\citenamefont {Gerra}\ \emph {et~al.}(2006)\citenamefont {Gerra},
  \citenamefont {Tagantsev}, \citenamefont {Setter},\ and\ \citenamefont
  {Parlinski}}]{Gerra:2006}%
  \BibitemOpen
  \bibfield  {author} {\bibinfo {author} {\bibfnamefont {G.}~\bibnamefont
  {Gerra}}, \bibinfo {author} {\bibfnamefont {A.~K.}\ \bibnamefont
  {Tagantsev}}, \bibinfo {author} {\bibfnamefont {N.}~\bibnamefont {Setter}}, \
  and\ \bibinfo {author} {\bibfnamefont {K.}~\bibnamefont {Parlinski}},\
  }\bibfield  {title} {\enquote {\bibinfo {title} {Ionic polarizability of
  conductive metal oxides and critical thickness for ferroelectricity in
  ${\mathrm{batio}}_{3}$},}\ }\href@noop {} {\bibfield  {journal} {\bibinfo
  {journal} {Phys. Rev. Lett.}\ }\textbf {\bibinfo {volume} {96}},\ \bibinfo
  {pages} {107603} (\bibinfo {year} {2006})}\BibitemShut {NoStop}%
\bibitem [{\citenamefont {Gerra}\ \emph {et~al.}(2007)\citenamefont {Gerra},
  \citenamefont {Tagantsev}, \citenamefont {Setter},\ and\ \citenamefont
  {Parlinski}}]{Gerra:2006v2}%
  \BibitemOpen
  \bibfield  {author} {\bibinfo {author} {\bibfnamefont {G.}~\bibnamefont
  {Gerra}}, \bibinfo {author} {\bibfnamefont {A.~K.}\ \bibnamefont
  {Tagantsev}}, \bibinfo {author} {\bibfnamefont {N.}~\bibnamefont {Setter}}, \
  and\ \bibinfo {author} {\bibfnamefont {K.}~\bibnamefont {Parlinski}},\
  }\bibfield  {title} {\enquote {\bibinfo {title} {Erratum: Ionic
  polarizability of conductive metal oxides and critical thickness for
  ferroelectricity in ${\mathrm{batio}}_{3}$ [phys. rev. lett. \textbf{96} ,
  107603 (2006)]},}\ }\href@noop {} {\bibfield  {journal} {\bibinfo  {journal}
  {Phys. Rev. Lett.}\ ,\ \bibinfo {pages} {169904}} (\bibinfo {year}
  {2007})}\BibitemShut {NoStop}%
\bibitem [{\citenamefont {Ishikawa}\ \emph {et~al.}(1996)\citenamefont
  {Ishikawa}, \citenamefont {Nomura}, \citenamefont {Okada},\ and\
  \citenamefont {Takada}}]{Ishikawa:1996}%
  \BibitemOpen
  \bibfield  {author} {\bibinfo {author} {\bibfnamefont {Kenji}\ \bibnamefont
  {Ishikawa}}, \bibinfo {author} {\bibfnamefont {Takashi}\ \bibnamefont
  {Nomura}}, \bibinfo {author} {\bibfnamefont {Nagaya}\ \bibnamefont {Okada}},
  \ and\ \bibinfo {author} {\bibfnamefont {Kazumasa}\ \bibnamefont {Takada}},\
  }\bibfield  {title} {\enquote {\bibinfo {title} {Size effect on the phase
  transition in p b t i o 3 fine particles},}\ }\href@noop {} {\bibfield
  {journal} {\bibinfo  {journal} {Japanese Journal of Applied Physics}\
  }\textbf {\bibinfo {volume} {35}},\ \bibinfo {pages} {5196} (\bibinfo {year}
  {1996})}\BibitemShut {NoStop}%
\bibitem [{\citenamefont {Tybell}\ \emph {et~al.}(1999)\citenamefont {Tybell},
  \citenamefont {Ahn},\ and\ \citenamefont {Triscone}}]{Tybell:1999}%
  \BibitemOpen
  \bibfield  {author} {\bibinfo {author} {\bibfnamefont {T.}~\bibnamefont
  {Tybell}}, \bibinfo {author} {\bibfnamefont {C.~H.}\ \bibnamefont {Ahn}}, \
  and\ \bibinfo {author} {\bibfnamefont {J.-M.}\ \bibnamefont {Triscone}},\
  }\bibfield  {title} {\enquote {\bibinfo {title} {Ferroelectricity in thin
  perovskite films},}\ }\href@noop {} {\bibfield  {journal} {\bibinfo
  {journal} {Applied Physics Letters}\ }\textbf {\bibinfo {volume} {75}},\
  \bibinfo {pages} {856--858} (\bibinfo {year} {1999})}\BibitemShut {NoStop}%
\bibitem [{\citenamefont {Jiang}\ \emph {et~al.}(2000)\citenamefont {Jiang},
  \citenamefont {Peng}, \citenamefont {Bursill},\ and\ \citenamefont
  {Zhong}}]{Jiang:2000}%
  \BibitemOpen
  \bibfield  {author} {\bibinfo {author} {\bibfnamefont {B.}~\bibnamefont
  {Jiang}}, \bibinfo {author} {\bibfnamefont {J.~L.}\ \bibnamefont {Peng}},
  \bibinfo {author} {\bibfnamefont {L.~A.}\ \bibnamefont {Bursill}}, \ and\
  \bibinfo {author} {\bibfnamefont {W.~L.}\ \bibnamefont {Zhong}},\ }\bibfield
  {title} {\enquote {\bibinfo {title} {Size effects on ferroelectricity of
  ultrafine particles of pbtio3},}\ }\href@noop {} {\bibfield  {journal}
  {\bibinfo  {journal} {Journal of Applied Physics}\ }\textbf {\bibinfo
  {volume} {87}},\ \bibinfo {pages} {3462--3467} (\bibinfo {year}
  {2000})}\BibitemShut {NoStop}%
\bibitem [{\citenamefont {Ghosez}\ and\ \citenamefont
  {Rabe}(2000)}]{Ghosezrabe:2000}%
  \BibitemOpen
  \bibfield  {author} {\bibinfo {author} {\bibfnamefont {Ph.}\ \bibnamefont
  {Ghosez}}\ and\ \bibinfo {author} {\bibfnamefont {K.~M.}\ \bibnamefont
  {Rabe}},\ }\bibfield  {title} {\enquote {\bibinfo {title} {Microscopic model
  of ferroelectricity in stress-free pbtio3 ultrathin films},}\ }\href@noop {}
  {\bibfield  {journal} {\bibinfo  {journal} {Applied Physics Letters}\
  }\textbf {\bibinfo {volume} {76}},\ \bibinfo {pages} {2767--2769} (\bibinfo
  {year} {2000})}\BibitemShut {NoStop}%
\bibitem [{\citenamefont {Meyer}\ and\ \citenamefont
  {Vanderbilt}(2001)}]{Meyer:2001}%
  \BibitemOpen
  \bibfield  {author} {\bibinfo {author} {\bibfnamefont {B.}~\bibnamefont
  {Meyer}}\ and\ \bibinfo {author} {\bibfnamefont {David}\ \bibnamefont
  {Vanderbilt}},\ }\bibfield  {title} {\enquote {\bibinfo {title} {\textit{Ab
  initio} study of ${\mathrm{batio}}_{3}$ and ${\mathrm{pbtio}}_{3}$ surfaces
  in external electric fields},}\ }\href@noop {} {\bibfield  {journal}
  {\bibinfo  {journal} {Phys. Rev. B}\ }\textbf {\bibinfo {volume} {63}},\
  \bibinfo {pages} {205426} (\bibinfo {year} {2001})}\BibitemShut {NoStop}%
\bibitem [{\citenamefont {Kornev}\ \emph {et~al.}(2004)\citenamefont {Kornev},
  \citenamefont {Fu},\ and\ \citenamefont {Bellaiche}}]{Kornev:2004}%
  \BibitemOpen
  \bibfield  {author} {\bibinfo {author} {\bibfnamefont {Igor}\ \bibnamefont
  {Kornev}}, \bibinfo {author} {\bibfnamefont {Huaxiang}\ \bibnamefont {Fu}}, \
  and\ \bibinfo {author} {\bibfnamefont {L.}~\bibnamefont {Bellaiche}},\
  }\bibfield  {title} {\enquote {\bibinfo {title} {Ultrathin films of
  ferroelectric solid solutions under a residual depolarizing field},}\
  }\href@noop {} {\bibfield  {journal} {\bibinfo  {journal} {Phys. Rev. Lett.}\
  }\textbf {\bibinfo {volume} {93}},\ \bibinfo {pages} {196104} (\bibinfo
  {year} {2004})}\BibitemShut {NoStop}%
\bibitem [{\citenamefont {Ramesh}\ and\ \citenamefont
  {Schlom}(2002)}]{RameshSchlom:2002}%
  \BibitemOpen
  \bibfield  {author} {\bibinfo {author} {\bibfnamefont {R.}~\bibnamefont
  {Ramesh}}\ and\ \bibinfo {author} {\bibfnamefont {D.~G.}\ \bibnamefont
  {Schlom}},\ }\bibfield  {title} {\enquote {\bibinfo {title} {{MATERIALS
  SCIENCE: Orienting Ferroelectric Films}},}\ }\href {\doibase
  10.1126/science.1072855} {\bibfield  {journal} {\bibinfo  {journal}
  {Science}\ }\textbf {\bibinfo {volume} {296}},\ \bibinfo {pages} {1975--1976}
  (\bibinfo {year} {2002})}\BibitemShut {NoStop}%
\bibitem [{\citenamefont {Despont}\ \emph {et~al.}(2006)\citenamefont
  {Despont}, \citenamefont {Koitzsch}, \citenamefont {Clerc}, \citenamefont
  {Garnier}, \citenamefont {Aebi}, \citenamefont {Lichtensteiger},
  \citenamefont {Triscone}, \citenamefont {Garcia~de Abajo}, \citenamefont
  {Bousquet},\ and\ \citenamefont {Ghosez}}]{Despont:2006}%
  \BibitemOpen
  \bibfield  {author} {\bibinfo {author} {\bibfnamefont {L.}~\bibnamefont
  {Despont}}, \bibinfo {author} {\bibfnamefont {C.}~\bibnamefont {Koitzsch}},
  \bibinfo {author} {\bibfnamefont {F.}~\bibnamefont {Clerc}}, \bibinfo
  {author} {\bibfnamefont {M.~G.}\ \bibnamefont {Garnier}}, \bibinfo {author}
  {\bibfnamefont {P.}~\bibnamefont {Aebi}}, \bibinfo {author} {\bibfnamefont
  {C.}~\bibnamefont {Lichtensteiger}}, \bibinfo {author} {\bibfnamefont
  {J.-M.}\ \bibnamefont {Triscone}}, \bibinfo {author} {\bibfnamefont {F.~J.}\
  \bibnamefont {Garcia~de Abajo}}, \bibinfo {author} {\bibfnamefont
  {E.}~\bibnamefont {Bousquet}}, \ and\ \bibinfo {author} {\bibfnamefont {Ph.}\
  \bibnamefont {Ghosez}},\ }\bibfield  {title} {\enquote {\bibinfo {title}
  {Direct evidence for ferroelectric polar distortion in ultrathin lead
  titanate perovskite films},}\ }\href {\doibase 10.1103/PhysRevB.73.094110}
  {\bibfield  {journal} {\bibinfo  {journal} {Phys. Rev. B}\ }\textbf {\bibinfo
  {volume} {73}},\ \bibinfo {pages} {094110} (\bibinfo {year}
  {2006})}\BibitemShut {NoStop}%
\bibitem [{\citenamefont {Zhang}\ \emph {et~al.}(2014)\citenamefont {Zhang},
  \citenamefont {Li}, \citenamefont {Shimada}, \citenamefont {Wang},\ and\
  \citenamefont {Kitamura}}]{Zhang:2014}%
  \BibitemOpen
  \bibfield  {author} {\bibinfo {author} {\bibfnamefont {Yajun}\ \bibnamefont
  {Zhang}}, \bibinfo {author} {\bibfnamefont {Gui-Ping}\ \bibnamefont {Li}},
  \bibinfo {author} {\bibfnamefont {Takahiro}\ \bibnamefont {Shimada}},
  \bibinfo {author} {\bibfnamefont {Jie}\ \bibnamefont {Wang}}, \ and\ \bibinfo
  {author} {\bibfnamefont {Takayuki}\ \bibnamefont {Kitamura}},\ }\bibfield
  {title} {\enquote {\bibinfo {title} {Disappearance of ferroelectric critical
  thickness in epitaxial ultrathin $\mathrm{BaZr}{\mathrm{o}}_{3}$ films},}\
  }\href@noop {} {\bibfield  {journal} {\bibinfo  {journal} {Phys. Rev. B}\
  }\textbf {\bibinfo {volume} {90}},\ \bibinfo {pages} {184107} (\bibinfo
  {year} {2014})}\BibitemShut {NoStop}%
\bibitem [{\citenamefont {Hirth}(2008)}]{Hirth:2008}%
  \BibitemOpen
  \bibfield  {author} {\bibinfo {author} {\bibfnamefont {John}\ \bibnamefont
  {Hirth}},\ }\href@noop {} {\emph {\bibinfo {title} {Dislocations in
  Solids}}}\ (\bibinfo  {publisher} {Elsevier},\ \bibinfo {year}
  {2008})\BibitemShut {NoStop}%
\bibitem [{\citenamefont {Garrity}\ \emph {et~al.}(2014)\citenamefont
  {Garrity}, \citenamefont {Rabe},\ and\ \citenamefont
  {Vanderbilt}}]{Garrity:2014}%
  \BibitemOpen
  \bibfield  {author} {\bibinfo {author} {\bibfnamefont {Kevin~F.}\
  \bibnamefont {Garrity}}, \bibinfo {author} {\bibfnamefont {Karin~M.}\
  \bibnamefont {Rabe}}, \ and\ \bibinfo {author} {\bibfnamefont {David}\
  \bibnamefont {Vanderbilt}},\ }\bibfield  {title} {\enquote {\bibinfo {title}
  {Hyperferroelectrics: Proper ferroelectrics with persistent polarization},}\
  }\href@noop {} {\bibfield  {journal} {\bibinfo  {journal} {Phys. Rev. Lett.}\
  }\textbf {\bibinfo {volume} {112}},\ \bibinfo {pages} {127601} (\bibinfo
  {year} {2014})}\BibitemShut {NoStop}%
\bibitem [{\citenamefont {Stengel}\ \emph {et~al.}(2012)\citenamefont
  {Stengel}, \citenamefont {Fennie},\ and\ \citenamefont
  {Ghosez}}]{Stengel:2012}%
  \BibitemOpen
  \bibfield  {author} {\bibinfo {author} {\bibfnamefont {Massimiliano}\
  \bibnamefont {Stengel}}, \bibinfo {author} {\bibfnamefont {Craig~J.}\
  \bibnamefont {Fennie}}, \ and\ \bibinfo {author} {\bibfnamefont {Philippe}\
  \bibnamefont {Ghosez}},\ }\bibfield  {title} {\enquote {\bibinfo {title}
  {Electrical properties of improper ferroelectrics from first principles},}\
  }\href@noop {} {\bibfield  {journal} {\bibinfo  {journal} {Phys. Rev. B}\
  }\textbf {\bibinfo {volume} {86}},\ \bibinfo {pages} {094112} (\bibinfo
  {year} {2012})}\BibitemShut {NoStop}%
\bibitem [{\citenamefont {Sai}\ \emph {et~al.}(2009)\citenamefont {Sai},
  \citenamefont {Fennie},\ and\ \citenamefont
  {Demkov}}]{PhysRevLett.102.107601}%
  \BibitemOpen
  \bibfield  {author} {\bibinfo {author} {\bibfnamefont {Na}~\bibnamefont
  {Sai}}, \bibinfo {author} {\bibfnamefont {Craig~J.}\ \bibnamefont {Fennie}},
  \ and\ \bibinfo {author} {\bibfnamefont {Alexander~A.}\ \bibnamefont
  {Demkov}},\ }\bibfield  {title} {\enquote {\bibinfo {title} {Absence of
  critical thickness in an ultrathin improper ferroelectric film},}\ }\href
  {\doibase 10.1103/PhysRevLett.102.107601} {\bibfield  {journal} {\bibinfo
  {journal} {Phys. Rev. Lett.}\ }\textbf {\bibinfo {volume} {102}},\ \bibinfo
  {pages} {107601} (\bibinfo {year} {2009})}\BibitemShut {NoStop}%
\bibitem [{\citenamefont {Bennett}\ \emph {et~al.}(2012)\citenamefont
  {Bennett}, \citenamefont {Garrity}, \citenamefont {Rabe},\ and\ \citenamefont
  {Vanderbilt}}]{Bennett:2012}%
  \BibitemOpen
  \bibfield  {author} {\bibinfo {author} {\bibfnamefont {Joseph~W.}\
  \bibnamefont {Bennett}}, \bibinfo {author} {\bibfnamefont {Kevin~F.}\
  \bibnamefont {Garrity}}, \bibinfo {author} {\bibfnamefont {Karin~M.}\
  \bibnamefont {Rabe}}, \ and\ \bibinfo {author} {\bibfnamefont {David}\
  \bibnamefont {Vanderbilt}},\ }\bibfield  {title} {\enquote {\bibinfo {title}
  {Hexagonal $abc$ semiconductors as ferroelectrics},}\ }\href@noop {}
  {\bibfield  {journal} {\bibinfo  {journal} {Phys. Rev. Lett.}\ }\textbf
  {\bibinfo {volume} {109}},\ \bibinfo {pages} {167602} (\bibinfo {year}
  {2012})}\BibitemShut {NoStop}%
\bibitem [{\citenamefont {Stengel}\ and\ \citenamefont
  {Spaldin}(2006)}]{Stengel/Spaldin:2006}%
  \BibitemOpen
  \bibfield  {author} {\bibinfo {author} {\bibfnamefont {M.}~\bibnamefont
  {Stengel}}\ and\ \bibinfo {author} {\bibfnamefont {N.~A.}\ \bibnamefont
  {Spaldin}},\ }\bibfield  {title} {\enquote {\bibinfo {title} {Origin of the
  dielectric dead layer in nanoscale capacitors},}\ }\href@noop {} {\bibfield
  {journal} {\bibinfo  {journal} {Nature}\ }\textbf {\bibinfo {volume} {443}},\
  \bibinfo {pages} {679--682} (\bibinfo {year} {2006})}\BibitemShut {NoStop}%
\bibitem [{\citenamefont {Stengel}\ \emph {et~al.}(2009)\citenamefont
  {Stengel}, \citenamefont {Vanderbilt},\ and\ \citenamefont
  {Spaldin}}]{Stengel/Vanderbilt/Spaldin:2009a}%
  \BibitemOpen
  \bibfield  {author} {\bibinfo {author} {\bibfnamefont {Massimiliano}\
  \bibnamefont {Stengel}}, \bibinfo {author} {\bibfnamefont {David}\
  \bibnamefont {Vanderbilt}}, \ and\ \bibinfo {author} {\bibfnamefont
  {Nicola~A.}\ \bibnamefont {Spaldin}},\ }\bibfield  {title} {\enquote
  {\bibinfo {title} {Enhancement of ferroelectricity at metal-oxide
  interfaces},}\ }\href@noop {} {\bibfield  {journal} {\bibinfo  {journal}
  {Nature Materials}\ }\textbf {\bibinfo {volume} {8}},\ \bibinfo {pages}
  {392--397} (\bibinfo {year} {2009})}\BibitemShut {NoStop}%
\bibitem [{\citenamefont {Cai}\ \emph {et~al.}(2011)\citenamefont {Cai},
  \citenamefont {Zheng}, \citenamefont {Ma},\ and\ \citenamefont
  {Woo}}]{Caimeng:2011}%
  \BibitemOpen
  \bibfield  {author} {\bibinfo {author} {\bibfnamefont {Meng-Qiu}\
  \bibnamefont {Cai}}, \bibinfo {author} {\bibfnamefont {Yue}\ \bibnamefont
  {Zheng}}, \bibinfo {author} {\bibfnamefont {Pui-Wai}\ \bibnamefont {Ma}}, \
  and\ \bibinfo {author} {\bibfnamefont {C.~H.}\ \bibnamefont {Woo}},\
  }\bibfield  {title} {\enquote {\bibinfo {title} {Vanishing critical thickness
  in asymmetric ferroelectric tunnel junctions: First principle simulations},}\
  }\href@noop {} {\bibfield  {journal} {\bibinfo  {journal} {Journal of Applied
  Physics}\ }\textbf {\bibinfo {volume} {109}},\ \bibinfo {eid} {024103}
  (\bibinfo {year} {2011})}\BibitemShut {NoStop}%
\bibitem [{\citenamefont {Yamada}\ \emph {et~al.}(2015)\citenamefont {Yamada},
  \citenamefont {Tsurumaki-Fukuchi}, \citenamefont {Kobayashi}, \citenamefont
  {Nagai}, \citenamefont {Toyosaki}, \citenamefont {Kumigashira},\ and\
  \citenamefont {Sawa}}]{ADFM:ADFM201500371}%
  \BibitemOpen
  \bibfield  {author} {\bibinfo {author} {\bibfnamefont {Hiroyuki}\
  \bibnamefont {Yamada}}, \bibinfo {author} {\bibfnamefont {Atsushi}\
  \bibnamefont {Tsurumaki-Fukuchi}}, \bibinfo {author} {\bibfnamefont {Masaki}\
  \bibnamefont {Kobayashi}}, \bibinfo {author} {\bibfnamefont {Takuro}\
  \bibnamefont {Nagai}}, \bibinfo {author} {\bibfnamefont {Yoshikiyo}\
  \bibnamefont {Toyosaki}}, \bibinfo {author} {\bibfnamefont {Hiroshi}\
  \bibnamefont {Kumigashira}}, \ and\ \bibinfo {author} {\bibfnamefont
  {Akihito}\ \bibnamefont {Sawa}},\ }\bibfield  {title} {\enquote {\bibinfo
  {title} {{Strong Surface-Termination Effect on Electroresistance in
  Ferroelectric Tunnel Junctions}},}\ }\href {\doibase 10.1002/adfm.201500371}
  {\bibfield  {journal} {\bibinfo  {journal} {Advanced Functional Materials}\
  }\textbf {\bibinfo {volume} {25}},\ \bibinfo {pages} {2708--2714} (\bibinfo
  {year} {2015})}\BibitemShut {NoStop}%
\bibitem [{\citenamefont {Shi}\ \emph {et~al.}(2013)\citenamefont {Shi},
  \citenamefont {Guo}, \citenamefont {Wang}, \citenamefont {Princep},
  \citenamefont {Khalyavin}, \citenamefont {Manuel}, \citenamefont {Michiue},
  \citenamefont {Sato}, \citenamefont {Tsuda}, \citenamefont {Yu},
  \citenamefont {Arai}, \citenamefont {Shirako}, \citenamefont {Akaogi},
  \citenamefont {Wang}, \citenamefont {Yamaura},\ and\ \citenamefont
  {Boothroyd}}]{Shi:2013}%
  \BibitemOpen
  \bibfield  {author} {\bibinfo {author} {\bibfnamefont {Youguo}\ \bibnamefont
  {Shi}}, \bibinfo {author} {\bibfnamefont {Yanfeng}\ \bibnamefont {Guo}},
  \bibinfo {author} {\bibfnamefont {Xia}\ \bibnamefont {Wang}}, \bibinfo
  {author} {\bibfnamefont {Andrew~J.}\ \bibnamefont {Princep}}, \bibinfo
  {author} {\bibfnamefont {Dmitry}\ \bibnamefont {Khalyavin}}, \bibinfo
  {author} {\bibfnamefont {Pascal}\ \bibnamefont {Manuel}}, \bibinfo {author}
  {\bibfnamefont {Yuichi}\ \bibnamefont {Michiue}}, \bibinfo {author}
  {\bibfnamefont {Akira}\ \bibnamefont {Sato}}, \bibinfo {author}
  {\bibfnamefont {Kenji}\ \bibnamefont {Tsuda}}, \bibinfo {author}
  {\bibfnamefont {Shan}\ \bibnamefont {Yu}}, \bibinfo {author} {\bibfnamefont
  {Masao}\ \bibnamefont {Arai}}, \bibinfo {author} {\bibfnamefont {Yuichi}\
  \bibnamefont {Shirako}}, \bibinfo {author} {\bibfnamefont {Masaki}\
  \bibnamefont {Akaogi}}, \bibinfo {author} {\bibfnamefont {Nanlin}\
  \bibnamefont {Wang}}, \bibinfo {author} {\bibfnamefont {Kazunari}\
  \bibnamefont {Yamaura}}, \ and\ \bibinfo {author} {\bibfnamefont {Andrew~T.}\
  \bibnamefont {Boothroyd}},\ }\bibfield  {title} {\enquote {\bibinfo {title}
  {A ferroelectric-like structural transition in a metal},}\ }\href@noop {}
  {\bibfield  {journal} {\bibinfo  {journal} {Nature Materials}\ }\textbf
  {\bibinfo {volume} {12}},\ \bibinfo {pages} {1024} (\bibinfo {year}
  {2013})}\BibitemShut {NoStop}%
\bibitem [{\citenamefont {Kim}\ \emph {et~al.}(2016)\citenamefont {Kim},
  \citenamefont {Puggioni}, \citenamefont {Yuan}, \citenamefont {L.~Xie},
  \citenamefont {Campbell}, \citenamefont {Ryan}, \citenamefont {Choi},
  \citenamefont {Kim}, \citenamefont {Patzner}, \citenamefont {Ryu},
  \citenamefont {Podkaminer}, \citenamefont {Irwin}, \citenamefont {Ma},
  \citenamefont {Fennie}, \citenamefont {Rzchowski}, \citenamefont {X.~Q.~Pan},
  \citenamefont {Rondinelli},\ and\ \citenamefont {Eom}}]{Kim:2016}%
  \BibitemOpen
  \bibfield  {author} {\bibinfo {author} {\bibfnamefont {T.~H.}\ \bibnamefont
  {Kim}}, \bibinfo {author} {\bibfnamefont {D.}~\bibnamefont {Puggioni}},
  \bibinfo {author} {\bibfnamefont {Y.}~\bibnamefont {Yuan}}, \bibinfo {author}
  {\bibfnamefont {H.~Zhou}\ \bibnamefont {L.~Xie}}, \bibinfo {author}
  {\bibfnamefont {N.}~\bibnamefont {Campbell}}, \bibinfo {author}
  {\bibfnamefont {P.~J.}\ \bibnamefont {Ryan}}, \bibinfo {author}
  {\bibfnamefont {Y.}~\bibnamefont {Choi}}, \bibinfo {author} {\bibfnamefont
  {J.-W.}\ \bibnamefont {Kim}}, \bibinfo {author} {\bibfnamefont {J.~R.}\
  \bibnamefont {Patzner}}, \bibinfo {author} {\bibfnamefont {S.}~\bibnamefont
  {Ryu}}, \bibinfo {author} {\bibfnamefont {J.~P.}\ \bibnamefont {Podkaminer}},
  \bibinfo {author} {\bibfnamefont {J.}~\bibnamefont {Irwin}}, \bibinfo
  {author} {\bibfnamefont {Y.}~\bibnamefont {Ma}}, \bibinfo {author}
  {\bibfnamefont {C.~J.}\ \bibnamefont {Fennie}}, \bibinfo {author}
  {\bibfnamefont {M.~S.}\ \bibnamefont {Rzchowski}}, \bibinfo {author}
  {\bibfnamefont {V.~Gopalan}\ \bibnamefont {X.~Q.~Pan}}, \bibinfo {author}
  {\bibfnamefont {J.~M.}\ \bibnamefont {Rondinelli}}, \ and\ \bibinfo {author}
  {\bibfnamefont {C.~B.}\ \bibnamefont {Eom}},\ }\bibfield  {title} {\enquote
  {\bibinfo {title} {Polar metals by geometric design},}\ }\href@noop {}
  {\bibfield  {journal} {\bibinfo  {journal} {Nature}\ }\textbf {\bibinfo
  {volume} {533}},\ \bibinfo {pages} {68--72} (\bibinfo {year}
  {2016})}\BibitemShut {NoStop}%
\bibitem [{\citenamefont {Puggioni}\ and\ \citenamefont
  {Rondinelli}(2013)}]{Puggioni:2013}%
  \BibitemOpen
  \bibfield  {author} {\bibinfo {author} {\bibfnamefont {Danilo}\ \bibnamefont
  {Puggioni}}\ and\ \bibinfo {author} {\bibfnamefont {James~M.}\ \bibnamefont
  {Rondinelli}},\ }\bibfield  {title} {\enquote {\bibinfo {title} {Designing a
  robustly metallic noncentrosymmetric ruthenate oxide with large thermopower
  anisotropy},}\ }\href {http://dx.doi.org/10.1038/ncomms4432} {\bibfield
  {journal} {\bibinfo  {journal} {Nat. Commun.}\ }\textbf {\bibinfo {volume}
  {5}},\ \bibinfo {pages} {3432} (\bibinfo {year} {2013})}\BibitemShut
  {NoStop}%
\bibitem [{\citenamefont {Heyd}\ \emph {et~al.}(2003)\citenamefont {Heyd},
  \citenamefont {Scuseria},\ and\ \citenamefont {Ernzerhof}}]{Heyd2003}%
  \BibitemOpen
  \bibfield  {author} {\bibinfo {author} {\bibfnamefont {Jochen}\ \bibnamefont
  {Heyd}}, \bibinfo {author} {\bibfnamefont {Gustavo~E.}\ \bibnamefont
  {Scuseria}}, \ and\ \bibinfo {author} {\bibfnamefont {Matthias}\ \bibnamefont
  {Ernzerhof}},\ }\bibfield  {title} {\enquote {\bibinfo {title} {Hybrid
  functionals based on a screened coulomb potential},}\ }\href {\doibase
  http://dx.doi.org/10.1063/1.1564060} {\bibfield  {journal} {\bibinfo
  {journal} {J. Chem. Phys.}\ }\textbf {\bibinfo {volume} {118}},\ \bibinfo
  {pages} {8207--8215} (\bibinfo {year} {2003})}\BibitemShut {NoStop}%
\bibitem [{\citenamefont {Heyd}\ \emph {et~al.}(2006)\citenamefont {Heyd},
  \citenamefont {Scuseria},\ and\ \citenamefont {Ernzerhof}}]{Heyd2006}%
  \BibitemOpen
  \bibfield  {author} {\bibinfo {author} {\bibfnamefont {Jochen}\ \bibnamefont
  {Heyd}}, \bibinfo {author} {\bibfnamefont {Gustavo~E.}\ \bibnamefont
  {Scuseria}}, \ and\ \bibinfo {author} {\bibfnamefont {Matthias}\ \bibnamefont
  {Ernzerhof}},\ }\bibfield  {title} {\enquote {\bibinfo {title} {Erratum:
  “hybrid functionals based on a screened coulomb potential” [j. chem.
  phys.118, 8207 (2003)]},}\ }\href {\doibase
  http://dx.doi.org/10.1063/1.2204597} {\bibfield  {journal} {\bibinfo
  {journal} {J. Chem. Phys.}\ }\textbf {\bibinfo {volume} {124}},\ \bibinfo
  {pages} {219906} (\bibinfo {year} {2006})}\BibitemShut {NoStop}%
\bibitem [{\citenamefont {Kresse}\ and\ \citenamefont
  {Furthm\"uller}(1996)}]{Kresse/Furthmuller:1996b}%
  \BibitemOpen
  \bibfield  {author} {\bibinfo {author} {\bibfnamefont {G.}~\bibnamefont
  {Kresse}}\ and\ \bibinfo {author} {\bibfnamefont {J.}~\bibnamefont
  {Furthm\"uller}},\ }\bibfield  {title} {\enquote {\bibinfo {title}
  {Efficiency of ab-initio total energy calculations for metals and
  semiconductors using a plane-wave basis set},}\ }\href@noop {} {\bibfield
  {journal} {\bibinfo  {journal} {Computational Materials Science}\ }\textbf
  {\bibinfo {volume} {6}},\ \bibinfo {pages} {15 -- 50} (\bibinfo {year}
  {1996})}\BibitemShut {NoStop}%
\bibitem [{\citenamefont {Bl\"ochl}\ \emph {et~al.}(1994)\citenamefont
  {Bl\"ochl}, \citenamefont {Jepsen},\ and\ \citenamefont
  {Andersen}}]{Blochl/Jepsen/Andersen:1994}%
  \BibitemOpen
  \bibfield  {author} {\bibinfo {author} {\bibfnamefont {Peter~E.}\
  \bibnamefont {Bl\"ochl}}, \bibinfo {author} {\bibfnamefont {O.}~\bibnamefont
  {Jepsen}}, \ and\ \bibinfo {author} {\bibfnamefont {O.~K.}\ \bibnamefont
  {Andersen}},\ }\bibfield  {title} {\enquote {\bibinfo {title} {Improved
  tetrahedron method for brillouin-zone integrations},}\ }\href@noop {}
  {\bibfield  {journal} {\bibinfo  {journal} {Physical Review B}\ }\textbf
  {\bibinfo {volume} {49}},\ \bibinfo {pages} {16223--16233} (\bibinfo {year}
  {1994})}\BibitemShut {NoStop}%
\bibitem [{\citenamefont {Monkhorst}\ and\ \citenamefont
  {Pack}(1976)}]{Monkhorst/Pack:1976}%
  \BibitemOpen
  \bibfield  {author} {\bibinfo {author} {\bibfnamefont {Hendrik~J.}\
  \bibnamefont {Monkhorst}}\ and\ \bibinfo {author} {\bibfnamefont {James~D.}\
  \bibnamefont {Pack}},\ }\bibfield  {title} {\enquote {\bibinfo {title}
  {{Special points for Brillouin-zone integrations}},}\ }\href@noop {}
  {\bibfield  {journal} {\bibinfo  {journal} {Physical Review B}\ }\textbf
  {\bibinfo {volume} {13}},\ \bibinfo {pages} {5188--5192} (\bibinfo {year}
  {1976})}\BibitemShut {NoStop}%
\bibitem [{\citenamefont {P.}\ and\ \citenamefont {W}(1976)}]{Zeitschrift}%
  \BibitemOpen
  \bibfield  {author} {\bibinfo {author} {\bibfnamefont {Seidel}\ \bibnamefont
  {P.}}\ and\ \bibinfo {author} {\bibfnamefont {Hoffmann}\ \bibnamefont {W}},\
  }\bibfield  {title} {\enquote {\bibinfo {title} {Verfeinerung der
  kristallstruktur von {NaNbO$_3$}},}\ }\href@noop {} {\bibfield  {journal}
  {\bibinfo  {journal} {Zeitschrift fur Kristallographie}\ }\textbf {\bibinfo
  {volume} {143}},\ \bibinfo {pages} {444--459} (\bibinfo {year}
  {1976})}\BibitemShut {NoStop}%
\bibitem [{\citenamefont {Liu}\ \emph {et~al.}(2013)\citenamefont {Liu},
  \citenamefont {Kargarian}, \citenamefont {Kareev}, \citenamefont {Gray},
  \citenamefont {Ryan}, \citenamefont {Cruz}, \citenamefont {Tahir},
  \citenamefont {Chuang}, \citenamefont {Guo}, \citenamefont {Rondinelli},
  \citenamefont {Freeland}, \citenamefont {Fiete},\ and\ \citenamefont
  {Chakhalian}}]{Liu:2013}%
  \BibitemOpen
  \bibfield  {author} {\bibinfo {author} {\bibfnamefont {J.}~\bibnamefont
  {Liu}}, \bibinfo {author} {\bibfnamefont {M.}~\bibnamefont {Kargarian}},
  \bibinfo {author} {\bibfnamefont {M.}~\bibnamefont {Kareev}}, \bibinfo
  {author} {\bibfnamefont {B.}~\bibnamefont {Gray}}, \bibinfo {author}
  {\bibfnamefont {Phil~J.}\ \bibnamefont {Ryan}}, \bibinfo {author}
  {\bibfnamefont {A.}~\bibnamefont {Cruz}}, \bibinfo {author} {\bibfnamefont
  {N.}~\bibnamefont {Tahir}}, \bibinfo {author} {\bibfnamefont {Yi-De}\
  \bibnamefont {Chuang}}, \bibinfo {author} {\bibfnamefont {J.}~\bibnamefont
  {Guo}}, \bibinfo {author} {\bibfnamefont {James~M.}\ \bibnamefont
  {Rondinelli}}, \bibinfo {author} {\bibfnamefont {John~W.}\ \bibnamefont
  {Freeland}}, \bibinfo {author} {\bibfnamefont {Gregory~A.}\ \bibnamefont
  {Fiete}}, \ and\ \bibinfo {author} {\bibfnamefont {J.}~\bibnamefont
  {Chakhalian}},\ }\bibfield  {title} {\enquote {\bibinfo {title}
  {Heterointerface engineered electronic and magnetic phases of {NdNiO$_3$}
  thin films},}\ }\href@noop {} {\bibfield  {journal} {\bibinfo  {journal}
  {Nat. Commun.}\ }\textbf {\bibinfo {volume} {4}},\ \bibinfo {pages} {3714}
  (\bibinfo {year} {2013})}\BibitemShut {NoStop}%
\bibitem [{\citenamefont {Campbell}\ \emph {et~al.}(2006)\citenamefont
  {Campbell}, \citenamefont {Stokes}, \citenamefont {Tanner},\ and\
  \citenamefont {Hatch}}]{Isodisplace:2006}%
  \BibitemOpen
  \bibfield  {author} {\bibinfo {author} {\bibfnamefont {Branton~J.}\
  \bibnamefont {Campbell}}, \bibinfo {author} {\bibfnamefont {Harold~T.}\
  \bibnamefont {Stokes}}, \bibinfo {author} {\bibfnamefont {David~E.}\
  \bibnamefont {Tanner}}, \ and\ \bibinfo {author} {\bibfnamefont {Dorian~M.}\
  \bibnamefont {Hatch}},\ }\bibfield  {title} {\enquote {\bibinfo {title}
  {{{\it ISODISPLACE}: a web-based tool for exploring structural
  distortions}},}\ }\href@noop {} {\bibfield  {journal} {\bibinfo  {journal}
  {Journal of Applied Crystallography}\ }\textbf {\bibinfo {volume} {39}},\
  \bibinfo {pages} {607--614} (\bibinfo {year} {2006})}\BibitemShut {NoStop}%
\bibitem [{\citenamefont {Giovannetti}\ and\ \citenamefont
  {Capone}(2014)}]{PhysRevB.90.195113}%
  \BibitemOpen
  \bibfield  {author} {\bibinfo {author} {\bibfnamefont {Gianluca}\
  \bibnamefont {Giovannetti}}\ and\ \bibinfo {author} {\bibfnamefont {Massimo}\
  \bibnamefont {Capone}},\ }\bibfield  {title} {\enquote {\bibinfo {title}
  {Dual nature of the ferroelectric and metallic state in
  ${\mathrm{lioso}}_{3}$},}\ }\href {\doibase 10.1103/PhysRevB.90.195113}
  {\bibfield  {journal} {\bibinfo  {journal} {Phys. Rev. B}\ }\textbf {\bibinfo
  {volume} {90}},\ \bibinfo {pages} {195113} (\bibinfo {year}
  {2014})}\BibitemShut {NoStop}%
\bibitem [{\citenamefont {Zhong}\ \emph {et~al.}(1994)\citenamefont {Zhong},
  \citenamefont {King-Smith},\ and\ \citenamefont
  {Vanderbilt}}]{PhysRevLett.72.3618}%
  \BibitemOpen
  \bibfield  {author} {\bibinfo {author} {\bibfnamefont {W.}~\bibnamefont
  {Zhong}}, \bibinfo {author} {\bibfnamefont {R.~D.}\ \bibnamefont
  {King-Smith}}, \ and\ \bibinfo {author} {\bibfnamefont {David}\ \bibnamefont
  {Vanderbilt}},\ }\bibfield  {title} {\enquote {\bibinfo {title} {Giant lo-to
  splittings in perovskite ferroelectrics},}\ }\href {\doibase
  10.1103/PhysRevLett.72.3618} {\bibfield  {journal} {\bibinfo  {journal}
  {Phys. Rev. Lett.}\ }\textbf {\bibinfo {volume} {72}},\ \bibinfo {pages}
  {3618--3621} (\bibinfo {year} {1994})}\BibitemShut {NoStop}%
\bibitem [{\citenamefont {Mishra}\ \emph {et~al.}(2007)\citenamefont {Mishra},
  \citenamefont {Choudhury}, \citenamefont {Chaplot}, \citenamefont {Krishna},\
  and\ \citenamefont {Mittal}}]{PhysRevB.76.024110}%
  \BibitemOpen
  \bibfield  {author} {\bibinfo {author} {\bibfnamefont {S.~K.}\ \bibnamefont
  {Mishra}}, \bibinfo {author} {\bibfnamefont {N.}~\bibnamefont {Choudhury}},
  \bibinfo {author} {\bibfnamefont {S.~L.}\ \bibnamefont {Chaplot}}, \bibinfo
  {author} {\bibfnamefont {P.~S.~R.}\ \bibnamefont {Krishna}}, \ and\ \bibinfo
  {author} {\bibfnamefont {R.}~\bibnamefont {Mittal}},\ }\bibfield  {title}
  {\enquote {\bibinfo {title} {Competing antiferroelectric and ferroelectric
  interactions in $\mathrm{Na}\mathrm{Nb}{\mathrm{o}}_{3}$: Neutron diffraction
  and theoretical studies},}\ }\href {\doibase 10.1103/PhysRevB.76.024110}
  {\bibfield  {journal} {\bibinfo  {journal} {Phys. Rev. B}\ }\textbf {\bibinfo
  {volume} {76}},\ \bibinfo {pages} {024110} (\bibinfo {year}
  {2007})}\BibitemShut {NoStop}%
\bibitem [{\citenamefont {Perdew}(1985)}]{Perdewgap:1985}%
  \BibitemOpen
  \bibfield  {author} {\bibinfo {author} {\bibfnamefont {John~P.}\ \bibnamefont
  {Perdew}},\ }\bibfield  {title} {\enquote {\bibinfo {title} {Density
  functional theory and the band gap problem},}\ }\href {\doibase
  10.1002/qua.560280846} {\bibfield  {journal} {\bibinfo  {journal}
  {International Journal of Quantum Chemistry}\ }\textbf {\bibinfo {volume}
  {28}},\ \bibinfo {pages} {497--523} (\bibinfo {year} {1985})}\BibitemShut
  {NoStop}%
\bibitem [{\citenamefont {Stengel}\ \emph {et~al.}(2011)\citenamefont
  {Stengel}, \citenamefont {Aguado-Puente}, \citenamefont {Spaldin},\ and\
  \citenamefont {Junquera}}]{stengelpuente:2011}%
  \BibitemOpen
  \bibfield  {author} {\bibinfo {author} {\bibfnamefont {Massimiliano}\
  \bibnamefont {Stengel}}, \bibinfo {author} {\bibfnamefont {Pablo}\
  \bibnamefont {Aguado-Puente}}, \bibinfo {author} {\bibfnamefont {Nicola~A.}\
  \bibnamefont {Spaldin}}, \ and\ \bibinfo {author} {\bibfnamefont {Javier}\
  \bibnamefont {Junquera}},\ }\bibfield  {title} {\enquote {\bibinfo {title}
  {Band alignment at metal/ferroelectric interfaces: Insights and artifacts
  from first principles},}\ }\href {\doibase 10.1103/PhysRevB.83.235112}
  {\bibfield  {journal} {\bibinfo  {journal} {Phys. Rev. B}\ }\textbf {\bibinfo
  {volume} {83}},\ \bibinfo {pages} {235112} (\bibinfo {year}
  {2011})}\BibitemShut {NoStop}%
\bibitem [{\citenamefont {Giovannetti}\ \emph {et~al.}(2016)\citenamefont
  {Giovannetti}, \citenamefont {Puggioni}, \citenamefont {Rondinelli},\ and\
  \citenamefont {Capone}}]{Giovannetti_etal:2016}%
  \BibitemOpen
  \bibfield  {author} {\bibinfo {author} {\bibfnamefont {Gianluca}\
  \bibnamefont {Giovannetti}}, \bibinfo {author} {\bibfnamefont {Danilo}\
  \bibnamefont {Puggioni}}, \bibinfo {author} {\bibfnamefont {James~M.}\
  \bibnamefont {Rondinelli}}, \ and\ \bibinfo {author} {\bibfnamefont
  {Massimo}\ \bibnamefont {Capone}},\ }\bibfield  {title} {\enquote {\bibinfo
  {title} {{Interplay between electron correlations and polar displacements in
  metallic ${\mathrm{SrEuMo}}_{2}{\mathrm{O}}_{6}$}},}\ }\href {\doibase
  10.1103/PhysRevB.93.115147} {\bibfield  {journal} {\bibinfo  {journal} {Phys.
  Rev. B}\ }\textbf {\bibinfo {volume} {93}},\ \bibinfo {pages} {115147}
  (\bibinfo {year} {2016})}\BibitemShut {NoStop}%
\bibitem [{\citenamefont {Xiang}(2014)}]{Xiang:2014}%
  \BibitemOpen
  \bibfield  {author} {\bibinfo {author} {\bibfnamefont {H.~J.}\ \bibnamefont
  {Xiang}},\ }\bibfield  {title} {\enquote {\bibinfo {title} {{Origin of polar
  distortion in LiNbO$_3$-type ``ferroelectric'' metals:Role of A-site
  instability and short-range interactions}},}\ }\href@noop {} {\bibfield
  {journal} {\bibinfo  {journal} {Phys. Rev. B}\ }\textbf {\bibinfo {volume}
  {90}},\ \bibinfo {pages} {094108} (\bibinfo {year} {2014})}\BibitemShut
  {NoStop}%
\bibitem [{\citenamefont {Nukala}\ \emph {et~al.}(2017)\citenamefont {Nukala},
  \citenamefont {Ren}, \citenamefont {Agarwal}, \citenamefont {Berger},
  \citenamefont {Liu}, \citenamefont {Johnson},\ and\ \citenamefont
  {Agarwal}}]{Nukala:2017}%
  \BibitemOpen
  \bibfield  {author} {\bibinfo {author} {\bibfnamefont {Pavan}\ \bibnamefont
  {Nukala}}, \bibinfo {author} {\bibfnamefont {Mingliang}\ \bibnamefont {Ren}},
  \bibinfo {author} {\bibfnamefont {Rahul}\ \bibnamefont {Agarwal}}, \bibinfo
  {author} {\bibfnamefont {Jacob}\ \bibnamefont {Berger}}, \bibinfo {author}
  {\bibfnamefont {Gerui}\ \bibnamefont {Liu}}, \bibinfo {author} {\bibfnamefont
  {A.~T.~Charlie}\ \bibnamefont {Johnson}}, \ and\ \bibinfo {author}
  {\bibfnamefont {Ritesh}\ \bibnamefont {Agarwal}},\ }\bibfield  {title}
  {\enquote {\bibinfo {title} {Inverting polar domains via electrical pulsing
  in metallic germanium telluride},}\ }\href@noop {} {\bibfield  {journal}
  {\bibinfo  {journal} {Nat. Commun.}\ }\textbf {\bibinfo {volume} {8}},\
  \bibinfo {pages} {15033} (\bibinfo {year} {2017})}\BibitemShut {NoStop}%
\bibitem [{\citenamefont {Benedek}\ and\ \citenamefont
  {Birol}(2016)}]{Benedek:2016}%
  \BibitemOpen
  \bibfield  {author} {\bibinfo {author} {\bibfnamefont {Nicole~A.}\
  \bibnamefont {Benedek}}\ and\ \bibinfo {author} {\bibfnamefont {Turan}\
  \bibnamefont {Birol}},\ }\bibfield  {title} {\enquote {\bibinfo {title}
  {{'}ferroelectric{'} metals reexamined: fundamental mechanisms and design
  considerations for new materials},}\ }\href@noop {} {\bibfield  {journal}
  {\bibinfo  {journal} {J. Mater. Chem. C}\ }\textbf {\bibinfo {volume} {4}},\
  \bibinfo {pages} {4000--4015} (\bibinfo {year} {2016})}\BibitemShut {NoStop}%
\end{thebibliography}%

\newpage


\end{document}